\documentstyle[epsfig]{l-aa}

\def\ltsima{$\; \buildrel < \over \sim \;$}
\def\simlt{\lower.5ex\hbox{\ltsima}}
\def\gtsima{$\; \buildrel > \over \sim \;$}
\def\simgt{\lower.5ex\hbox{\gtsima}}

\begin{document}

\def\ltsima{$\; \buildrel < \over \sim \;$}
\def\simlt{\lower.5ex\hbox{\ltsima}}
   \thesaurus{03(02.02.1;02.02.2;11.14.1)}   

\title{A toroidal black hole for the AGN phenomenon}
\author{
Fulvio Pompilio
\inst{1}
\and S. M. Harun-or-Rashid
\inst{2} 
\and Matts Roos
\inst{2}}
\institute{SISSA (ISAS),Via Beirut 2-4, I--34013 Trieste, Italy\\
\and Division of High Energy Physics, PL-9, FIN-00014 
University of Helsinki, Finland}

\date{Received 14.7.2000/ Accepted }

\offprints{pompilio@sissa.it}

\maketitle

\markboth{F.Pompilio et al.:
A toroidal black hole for the AGN phenomenon}{}

\begin{abstract}

A new approach to the study of the AGN phenomenon is proposed, in which the 
nucleus  activity is related to the metric of the inner massive black hole. The
possibility of a Toroidal Black Hole (TBH), in contrast to the usual
Spherical Black Hole (SBH), is discussed as a powerful tool in understanding
AGN related phenomena, such as the energetics, the production of jets and
the acceleration of particles, the shape of the magnetic field and the 
lifetime of nucleus activity.\\


\end{abstract}

\section{Introduction}

Huge energetics of Active Galactic Nuclei (AGN) are usually interpreted in
terms of accretion around a massive black hole (Frank, King \& Raine 1985). 
So far, the standard
assumption has been that the black hole is a spherical (SBH), static
(i.e. described by the Schwarzschild metric) or else rotating around its axis
(i.e. described by the Kerr metric) one. However, models of collapse leading 
to different topologies of the resulting object have been investigated
(Smith \& Mann 1997) and a mathematical description of their properties in
the framework of general relativity has been developed by several 
authors, e.g. Vanzo (1997), Brill, Louko \& Peldan (1997). Recently, a few 
hints about the 
possible connection between topological models of black holes and high-energy
astrophysical phenomena like AGNs have been claimed (Spivey 2000). More
specifically, the possibility points toward a Toroidal Black Hole (TBH), in
which the event horizon is topologically equivalent to a torus, to explain
the AGN observations.\\
In this paper, we consider such a situation and evaluate the influence of 
this topology in accounting for it.\\
First, we give a basic description of the conditions needed to obtain a TBH
and the possibilities of matching them. Second, the geometry of a TBH is
characterized through a dynamical treatment of particle motion around it.
Then, some hints about the shape and intensity of magnetic field are
provided, yielding the emission properties of such a structure; along
this route, a particles acceleration and jet production mechanism is proposed.
Last, an interpretation of the finite lifetime of an AGN is deduced from the
whole picture.

\section{The toroidal structure for a black hole.}

\subsection{The spacetime metric.}

A generalization of the black hole metrics can be given as follows
(Smith \& Mann 1997):
\[{ds}^{2}=-\left( V+b-\frac{2M}{R}\right) {dt}^{2}+\frac{{dr}^{2}}
{\left( V+b-\frac{2M}{R}\right)}\]
\begin{equation} 
+{R}^{2}[{d\theta }^{2}+c~{sinh}^{2}(\sqrt{a}\theta ){d\phi }^{2}],
\end{equation}
where $t$, $r$ are the time and radial coordinates, $\theta $ and $\phi $ are
coordinates on a two-surface of constant curvature and $V$ is a potential
term, whose meaning will be discussed later. The parameters $b$, $c$, $a$
fix the topology of the structure: in particular, if $b=-a=0$ ($b=-a$ 
following the solution of Einstein field equations in empty space) and $a
\rightarrow 0$, $c=+\frac{1}{a}$, then the topology is that of a torus and
the space-time is asymptotically anti-de Sitter (AdS). We recall that a
Schwarzschild metric for an asymptotically flat spacetime exhibits $b=0$ 
and +1 instead
of the potential term $V$. Therefore, such a configuration is bounded to the
presence of $V$. A standard expression for the potential would be of the 
form:
\begin{equation}
V=\frac{\Lambda}{3}{R}^{2},
\end{equation}
where $\Lambda $ is a negative constant in order to provide an AdS 
spacetime. However, this cannot be the cosmological constant, which is known
to have a positive value (Perlmutter et al. 1997, 1999; Riess et al. 1998;
Roos \& Harun-or-Rashid 2000) and which is a few orders of magnitude smaller
than needed to act as an effective AdS term.
That is why another physical reasoning must be found for $V$ and it must be
of astrophysical nature (e.g. gravitational potential of the surrounding
galaxy); anyhow, some ideas about its origin will be further given.

\subsection{The geodesics.}

A simplified picture of the gravitational potential of a rotating TBH can be
gained by the motion of particles around it. Actually, particle dynamics is
the best tool to clarify the physical phenomena happening around the hole. 
Such a task can be pursued by solving the geodesics equations (Lawden 1982):
\begin{equation}
\frac{d}{ds}\left( 2~{g}_{ri}\frac{d{x}^{i}}{ds}\right) 
-\frac{\partial {g}_{jk}}{\partial {x}^{r}}\frac{d{x}^{j}}{ds}
\frac{d{x}^{k}}{ds}=0,
\end{equation}
for the different components ${g}_{ri}$ of the metric tensor. The 
corresponding radial equation can be stated as follows:
\begin{equation}
\frac{dr}{dt}={\left[ {c}^{2}({k}^{2}-V)+\frac{2MG{c}^{2}}{r}
+\frac{2MG{h}^{2}}{{c}^{2}{r}^{3}}-{h}^{2}V\right] }^{1/2},
\end{equation}
where $h$ is the specific angular momentum along the $z$-axis and $k$ is a
constant of integration. Therefore, the energy conservation equation can be
derived:
\begin{equation}
{\left( \frac{dr}{dt}\right) }^{2}-\frac{2MG{c}^{2}}{r}-
\frac{2MG{h}^{2}}{{c}^{2}{r}^{3}}={c}^{2}{k}^{2}+V({k}^{2}-{h}^{2}),
\end{equation}
which can be compared with the SBH situation:
\begin{equation}
{\left( \frac{dr}{dt}\right) }^{2}+\frac{{h}^{2}}{{r}^{2}}-\frac{2MG{c}^{2}}
{r}-\frac{2MG{h}^{2}}{{c}^{2}{r}^{3}}=const.
\end{equation}
The basic difference is based on the centrifugal term $\frac{{h}^{2}}
{{r}^{2}}$: indeed, if $V$ has no radial dependence, then the TBH 
\textit{has no centrifugal barrier} and, in order to reach the black hole,
the accreting matter will just need to overwhelm a
certain threshold value independent of $r$ and fixed by the right-hand side
of the above equation. We stress that this trend is of fundamental relevance
for the amount of angular momentum which must be exchanged by the infalling 
gas before accreting.\\
\begin{figure}
\begin{center}
\epsfig{file=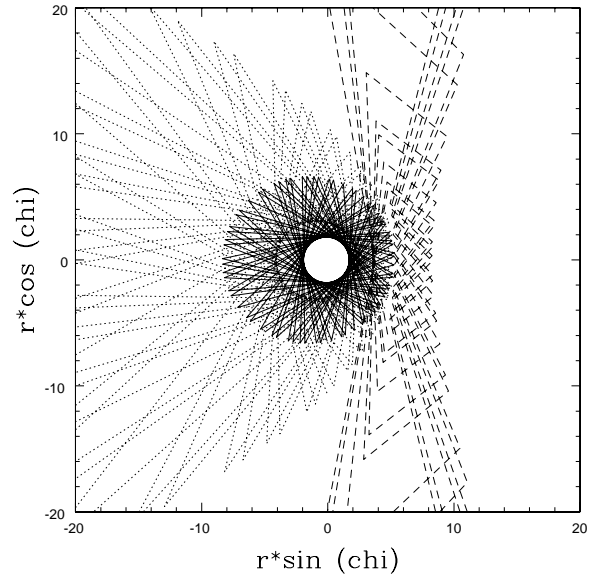,width=8cm}
\end{center}
\caption{Particle trajectories in the equatorial plane, assuming they are
experiencing just the TBH gravitational potential; the orbits are parametrized
as elliptical with $p=5.6$ and $V\simeq const$ is assumed. Different values
for the $e$ parameter have been assumed: (\textit{solid line}) $e=0.2$,
(\textit{short-dashed line}) $e=1.0$, (\textit{long-dashed line}) $e=4.5$.}
\end{figure}
To depict particle trajectories and in order to make a comparison with the 
SBH result, we followed the standard approach of 
parametrizing the orbit in a general elliptical way (Chandrasekhar 1983):
\begin{equation}
r(\chi )=\frac{p}{(1+e~cos\chi )},
\end{equation}
in terms of a dimensionless parameter $\chi $ along the orbit, then we solved
the time ($\frac{dt}{d\chi }$) and angular ($\frac{d\phi }{d\chi }$)
components of the geodesics, with zero initial conditions. In Fig.1 the 
trajectories in the equatorial plane are sketched for $V\simeq const$ 
with $p=5.6$ and different choices for the $e$ parameter. The 
geodesics have been evaluated for $e=0.2$, $e=1.0$ and $e=4.5$. We recall that,
in general, for a central potential of the form $V=-\frac{{V}_{0}}{r}$, 
it holds:
\begin{equation}
e=\sqrt{1+\frac{2{h}^{2}E}{{V}_{0}^{2}m}},
\end{equation}
for a particle with energy $E$ and mass $m$. Then, the first two choices for
the $e$ parameter correspond respectively to low and mildly energetic 
particles.
In this case, particles spiral around the torus and are focused in a circular
region internal to the structure, whose extent is wider for the more energetic
particles. As they turn around the torus they gain
acceleration and, in particular, while approaching the centre of symmetry
($r\rightarrow 0$), their acceleration is proportional to the time they spend
in the inner region, as follows:
\begin{equation}
\frac{dr}{dt}\sim \sqrt{2MG}\left( \frac{h}{c}\right) {r}^{-3/2}
\Rightarrow a\propto {t}^{-8/5},
\end{equation}
so that the less they spend in the center the more they can gain and we will
refer to this stage as a \textit{gravitational kick}: eventually, 
when a particle 
reaches the matching-condition orbit, it can be kicked out through the 
circular inner region. We note that $e=1.0$ provides a limiting
state, because a few particles can escape the black hole through a parabolic
orbit. Last, particles with $e$ greater than unity have too much energy to 
get trapped by the black hole, therefore they just approach it and experience 
a \textit{slingshot mechanism}, kicking them off.\\
If they are focused to the center, particles are likely to undergo collisions, 
which could be an
effective way to exchange angular momentum and can provide another mechanism
of acceleration by means of a Fermi-type stochastic procedure. In that
framework, the energy exchanged by a particle per unit time due to collisions
is proportional to the particle velocity $v$ and to the size $l$ of the 
acceleration region (Longair 1981):
\begin{equation}
\frac{dE}{dt}=\frac{4}{3}\left( \frac{{v}^{2}}{cl}\right)E. 
\end{equation}
The conservation equation depends on the the time spent by the particle
in that region ${t}_{f}$ and if we neglect diffusion and source terms, 
it holds:
\begin{equation}
\frac{dN}{dt}=-\frac{d}{dt}\left[ \alpha ~E~N(E)\right] -\frac{N}{{t}_{f}},
\end{equation}
where $\alpha \equiv \frac{{v}^{2}}{cl}$. The final spectrum of the particles
emitted by the core in a steady state situation ($\frac{dN}{dt}=0$) is given 
by:
\begin{equation}
N(E)\propto {E}^{-\left( 1+\frac{1}{\alpha {t}_{f}}\right) },
\end{equation}
so that \textit{a jet starting from the center will have a power-law 
distribution}.\\
We remind that to have an effective acceleration mechanism, the particles 
must have energies greater than the maximum energy loss felt during 
acceleration, or else the loss can be overcome through a sufficiently rapid 
initial acceleration mechanism: that is exactly what makes a TBH a likely 
candidate for an effective process.

\subsection{Extended particle dynamics.}

The spiral-shaped trajectory is not the only motion the particles undergo.
In addition to it the centrifugal and the Coriolis forces must be added;
moreover, if we consider the flow around the TBH to be embedded in the
magnetic field produced by the accretion disk, a Lorentz force term arises.
Actually, the accretion disk can be considered to be composed of annuli of
charged particles fluxes, providing azimuthal currents: this results in a 
magnetic field in which field lines encircle those currents. Therefore, the 
three forces can be evaluated as follows:
\begin{itemize}
\item ${F}_{cf}=\gamma {m}_{0}(\vec{\Omega}\times \vec{r})\times \vec{\Omega}$;
\item ${F}_{Cor}={m}_{0}\left[ 2\gamma \frac{dr}{dt}+r\frac{d\gamma }
{dt}\right] \left( \vec{{e}_{r}}\times \vec{\Omega} \right) $;
\item ${F}_{B}=\frac{q}{c}\left( \vec{v}\times \vec{B}\right) $;
\end{itemize}
for a particle of rest-mass ${m}_{0}$, charge $q$ and Lorentz factor 
$\gamma $. The effect of the Lorentz force is a drift in the azimuthal
coordinate, adding to the rounded spiral motion or eventually against it
and braking the flow; nevertheless, this motion provides a shear effect
and an energy exchange between $+/-$ charged flows. On the other hand, the
superposition of the centrifugal force and the Coriolis force (opposite to
the direction of rotation) can allow particles to move towards other orbits
and could be responsible for particles pointing to the centre of symmetry. 
The azimuthal drift develops a ring current associated to an encircling
magnetic field, which adds to the disk component. At the end, an axisymmetric
toroidal plasma configuration arises, resembling the \textit{tokamak}
device involved in nuclear fusion studies. The total magnetic field will
exhibit helical field lines (Dendy 1993):
\begin{equation}
\vec{B}=\frac{1}{R}\vec{\nabla }\psi \times \vec{{e}_{\phi }}+ {B}_{\phi }
\vec{{e}_{\phi }},
\end{equation}
where the potential $\psi $ can be inferred by means of the 
\textit{Grad-Shafranov equations}, derived by the pressure-balance condition
on the azimuthal component:
\begin{equation}
\left( \vec{j}\times \vec{B}\right) \cdot \vec{{e}_{\phi }}=\vec{\nabla }
p\cdot \vec{{e}_{\phi }}=0,
\end{equation}
and the radial component:
\begin{equation}
\left( \vec{j}\times \vec{B}\right) \cdot \vec{{e}_{r}}=\frac{\partial p}
{\partial r},
\end{equation}
and the plasma is self-confined to the tokamak shape by means of a 
\textit{$\theta$-pinch} and a \textit{$z$-pinch} effects.\\
However, it's worth stressing that this tokamak-like plasma structure has a
basic difference with respect to the real tokamak device, that is the strong
gravitational field. In fact, the plasma shape is mantained by means of the
TBH gravitation, in contrast to the external magnetic field involved in
tokamaks. Therefore, a strong gravitational force is likely to prevent 
diffusive motion, which appears in magnetic confined toroidal plasma. This
gives to the system enough stability to treat it by the Grad-Shafranov
equations.\\
In conclusion, particles spiral around the torus and eventually approach the
acceleration inner region, where they are kicked out by a gravitational
kick or they partake in the Fermi-type power-law
distribution and flow away along the TBH axis; there, they can be trapped by
the magnetic field, collimating them in a jet.\\
There is another process acting in such a configuration related to 
dynamical instabilities. Indeed, particle trajectories are supposed
to experience Lindblad resonances as follows from perturbations in the
gravitational potential (Binney \& Tremaine 1987). Both with particle 
collisions and energy/momentum exchange, instabilities can act as an
accretion mechanism, whenever the gravitational kick conditions are not 
matched. 
Gravitational instabilities could also influence the tokamak stability, but 
they should face plasma instabilities and the competition between the two 
fixes the final configuration.
Actually,
to be exhaustive and more quantitative, we should perform a 
hard magnetohydrodynamical (MHD) treatment 
of the particle motion. Since it requires a hard computational work, we 
cannot discuss it here; nevertheless, we recall that MHD 2D simulations
have shown that turbulence in the accretion disc has a fundamental role.
The matter in the region of unstable orbits does not follow simple energy
and angular momentum conserving free fall trajectories, but a deep exchange is
involved by means of turbulence on the way to the event horizon and also the
shear components are enormous (e.g. Hawley \& Krolik 2000, Igumenshchev \& 
Abramowicz 2000 and references therein). This will help putting matter onto
orbits which spiral around the torus, so that it easily reaches the inner
region, where it will experience the gravitational kick.

\section{The emission properties of a TBH.}

The huge wide-band emission from AGN is mostly explained through processes
involving the accretion disk (Frank, King \& Raine 1985). Nevertheless, 
the most energetic 
emission, i.e. gamma rays, is far from being self-consistently explained: the
observed gamma ray energy along the jets in quasars is close to four orders
of magnitude higher than expected by theory (Spivey 2000). In particular, the 
mechanism proposed by Blandford \& Znajek (1978) cannot account for such an
amount of energy, so that it cannot be the only responsible for it. As
discussed in the previous section, in the framework of a TBH a natural
explanation for jet production emerges, then it is important to
estimate the highest values for the Lorentz factor that particles can reach. 
The limit to that value is attained because of Inverse Compton
(IC) and synchrotron losses (Longair 1981):
\begin{itemize}
\item ${\gamma }_{max}^{IC}\simeq \sqrt{\frac{3\times {10}^{7}\omega }
{{U}_{rad}}}$;
\item ${\gamma }_{max}^{SYNC}\simeq \frac{\sqrt{5\times {10}^{8}\omega }}{B}$;
\end{itemize}
where ${U}_{rad}$ refers to the radiation field particles move in. It follows 
that in the approximation of a keplerian orbit at a distance corresponding
to the event horizon, the maximum Lorentz factor around a TBH is increased
by a factor:
\begin{equation}
\frac{{\gamma }_{max}^{TBH}}{{\gamma }_{max}^{SBH}}\propto 
\sqrt{\frac{{\omega}_{TBH}}{{\omega}_{SBH}}}\sim {V}^{3/4}, 
\end{equation}
for fixed values of the radiation field and the magnetic field.\\
As concerns the spectrum, a TBH can give rise to intense bremsstrahlung and 
synchrotron emission for a sufficiently high angular velocity $\omega $ 
attained by the spiralling particles. If we consider the aforementioned 
power-law distribution, then the emission due to relativistic bremsstrahlung,
synchrotron radiation and IC will be (Rybicki \& Lightman 1979):
\begin{itemize}
\item ${I}_{brems}\propto {\nu}^{-\left( 1+\frac{1}{\alpha {t}_{f}}\right) }$;
\item ${I}_{SYNC}\propto {B}^{\left( 1+\frac{1}{2\alpha {t}_{f}}\right)}
{\nu}^{-\left( \frac{1}{2\alpha {t}_{f}}\right) }$;
\item ${I}_{IC}\propto {{\nu }_{0}}^{\left( \frac{1}{2\alpha {t}_{f}}
-1\right) }{\nu}^{-\left( \frac{1}{2\alpha {t}_{f}}\right) }$;
\end{itemize}
where $B$ is the embedding magnetic field and ${\nu}_{0}$ is the injection
energy.\\
Moreover, the Blandford-Znajek mechanism is increased by the extra
magnetic field added to the embedding disc field, due to the dynamics leading
to the tokamak-like plasma configuration; as a result, the power extracted 
increases with respect to the SBH situation by a factor of 
(Frank, King \& Raine 1985):
\begin{equation}
{W}_{B-Z}\propto {\left( \frac{{B}_{TBH}}{{B}_{SBH}}\right) }^{2}.
\end{equation}
In addition, a new process could be involved in emission properties of a TBH,
due to the tokamak-like plasma, that is the
\textit{sawteeth emission}, i.e. an intense periodic burst arising from the
centre of a tokamak device, whose energy ranges from hard X-rays up to gamma
rays (Dendy 1993). Its physical origin is still debated, but it seems to 
depend on
magneto-hydrodynamical instabilities acting on the plasma, such that a slow
increase in plasma pressure is followed by an abrupt fall, the current rapidly
grows and magnetic reconnection feeds the burst.\\
Another emission process, that is purely related to the space-time 
characteristics, is the \textit{Penrose process} (Penrose 1969). 
It follows from the fact 
that in a rotating black hole the surface on which the time component of the
metric tensor vanishes, i.e. ${g}_{tt}=0$, differs from the event horizon:
the two surfaces enclose a region called \textit{ergosphere} where the energy 
of a particle as observed from far way can be negative. The original
\textit{Penrose process} follows from the decay of a particle into two
photons, one of which crosses the horizon and gets lost, the other one gains
energy in the range fixed by the Wald inequality (Wald 1984):
\begin{equation}
\gamma E-\gamma ~v{({E}^{2}-{g}_{tt})}^{1/2}\le \epsilon \le
\gamma E+\gamma ~v{({E}^{2}-{g}_{tt})}^{1/2},
\end{equation}
then a TBH is more efficient than a Kerr black hole in producing high-energy 
photons in this way, provided that ${\gamma }_{max}^{TBH}\ge 
{\gamma }_{max}^{SBH} $.

\section{Lifetime of AGN activity.}

If the AGN activity is related to the toroidal shape of the black hole, a
transition to a quiescent state is expected as the black hole reaches 
SBH status. In fact, as matter accretes the hole, the event horizon increases,
the torus inflates and lately looses its starting configuration turning
spherical. On the above line, we can infer a more quantitative estimate for
the lifetime of the activity phase by equating the metric tensor components
for a TBH and a SBH. If ${R}_{in}$ is an initial radial dimension for the 
torus, e.g. the middle value of the torus thickness with respect to the centre
of symmetry, and ${R}_{fin}$ is the final radius of the SBH, then the
transition happens if the following condition is matched:
\begin{equation}
V-\frac{2MG}{{R}_{in}{c}^{2}}=1-\frac{2MG}{{R}_{fin}{c}^{2}},
\end{equation}
or else:
\begin{equation}
{R}_{fin}=\frac{{r}_{g}}{1-V+\frac{{r}_{g}}{{R}_{in}}},
\end{equation}
where ${r}_{g}$ is the gravitational radius. Besides, the lifetime is related
to the accreted matter $\Delta M$ and to the accretion rate $\frac{dM}{dt}$
by:
\begin{equation}
dt(life)\sim \frac{\Delta M}{dM/dt},
\end{equation}
yielding:
\begin{equation}
dt(life)\propto \frac{1}{dM/dt}{{R}_{fin}}^{3}.
\end{equation}
Since $\frac{dM}{dt}$ is observationally estimated, 
we could enter the lifetime debate if only we knew the potential. 
Lacking that, we can only study $dt(life)$ as a function of some
hypothetical functional form for it:
\begin{enumerate}
\item $V\simeq const$;
\item $V\propto 1/R$;
\item $V\propto log~R$;
\end{enumerate}
where the first case could be regarded as a cosmological vacuum energy
density, while the other expressions could mimic the background
potential of a surrounding axisymmetric galaxy. In particular, the second
functional form refers to an embedding newtonian gravitational field;
on the contrary, the logarithmic shape is motivated in order to
reproduce the flatness of the galactic rotation curve (Binney \& 
Tremaine 1987). 
In this frame, the formation of a TBH can be deduced
if a protogalaxy develops a sufficient extended background potential and the
collapse to a massive black hole is forced to lead to a toroidal 
configuration\footnote{We suggest that another way to obtain a seed TBH should
follow the dynamical evolution of a black hole binary system; however, such a
topic deserves a strict treatment.}.\\
\begin{figure}
\begin{center}
\epsfig{file=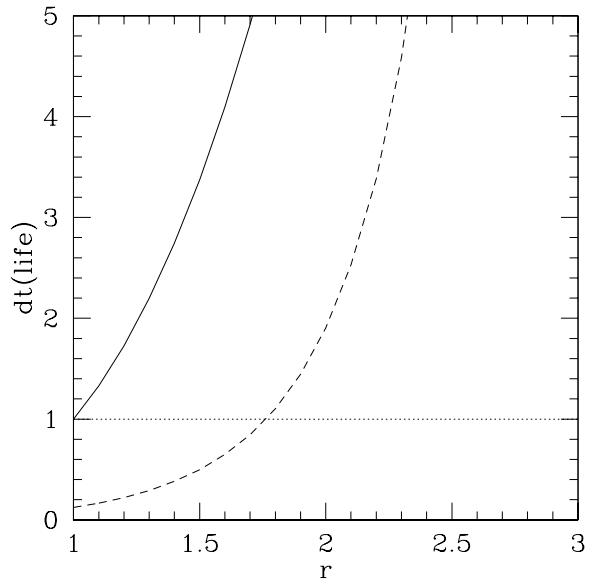,width=8cm}
\end{center}
\caption{AGN lifetime for a TBH model with constant accretion rate: 
(\textit{solid line}) $V\simeq const$; (\textit{short-dashed line}) 
$V\propto 1/R$; (\textit{long-dashed line}) $V\propto log~R$ .
The normalization has been arbitrarily fixed and $r\equiv \frac{{R}_{in}}
{{r}_{g}}$.}
\end{figure}
In Fig.2, the trend of $dt(life)$ is sketched for the above
mentioned different potentials, assuming a constant accretion rate $dM/dt$. 
The $1/R$ potential gives a fixed value independent of ${R}_{in}$, while the
$V\simeq const$ situation gives an increasing function, as the 
$log~R$ potential does. If the AGN lifetimes
have small dispersion and are strictly focused on a single value, e.g.
${10}^{6}-{10}^{7}$ yr (e.g. Cavaliere \& Vittorini 2000), then the $1/R$ 
potential seems to give a likely interpretation to it.\\
Last, we stress that the TBH/SBH transition is likely to be a violent
event in the galaxy evolution. From a physical point of view, the galactic
potential turns from being axisymmetric (TBH) and with a relevant external
component ($V$) to a spherical configuration dominated by the central SBH.
Therefore, a rapid and huge accretion should happen, the black hole swallows
the most of the surrounding, so that the activity stops and a quiescent state
is reached.

\section{Comparison between TBH and SBH.}

In this section, we trace a few guidelines to emphasize why a TBH could be
more convenient in explaining the AGN phenomenon than the standard SBH 
model.\\
First of all, the TBH provides a powerful mean to accelerate particles and to
give them high Lorentz factors\footnote{For sake of completeness, we stress
that such a mechanism could be efficient in other astrophysical contexts, e.g.
acceleration of Ultra Relativistic Cosmic Rays.}; on the contrary, a SBH has
more severe limitations in doing it and usually other complicated phenomena
involving the surrounding galaxy and inter-stellar medium must be invoked,
e.g. shocks induced by stellar ejections or hydrodynamical expansion (e.g.
Rees, Phinney, Begelman \& Blandford 1982).
Moreover, the variety of emission, which is well described by the standard SBH
model, is not restricted by a TBH model, which affects just the energetics. 
On the
other hand, our model could explain the production of jets in a more likely
way and could account for the large angle dispersion observed, 
as follows from the opening angle of the TBH.\\
An important new perspective with this model is the way it explains the
finite lifetime of the AGN phenomenon, that is hardly interpreted by the
standard scheme through a multiparameter modelling of black hole Initial
Mass Function (IMF), accretion rate and other quantities (e.g. Monaco, 
Salucci \& Danese 1999). Indeed, 
we stress the power of a TBH model in significantly simplifying the finite 
lifetime puzzle.\\
So far, we see no drawbacks for a TBH model with respect to a SBH model, but
we underline how it could improve and complete the general picture of the
AGN phenomenon. An argument which could be raised against our view could come
from iron emission line observations, well explained in terms of being
generated close to a spinning SBH, i.e. a Kerr black hole (e.g.
Fabian et al. 1989, Matt, Perola \& Stella 1993). Nevertheless, the emission
line arises several gravitational radii away from the hole, where particles
could not distinguish among a SBH and a TBH. However, the difference in the
relativistic gravitational shift effect should be:
\begin{equation}
\frac{\Delta \nu }{\nu }(TBH)=\frac{\Delta \nu }{\nu }(SBH){\left( 
\frac{V-3{r}_{g}/{R}}{1-{r}_{g}/{R}}\right) }^{1/2},
\end{equation}
and for $R\gg {r}_{g}$ it holds:
\begin{equation}
\frac{\Delta \nu }{\nu }(TBH)\propto \sqrt{V}\frac{\Delta \nu }{\nu }(SBH),
\end{equation}
therefore this effect is likely to be undistinguished by recent X-ray data;
besides, the deep model dependence of the Kerr black hole interpretation
of the iron line gives no hope for a choice among the two models, until 
higher resolution observation will be available by the next X-ray satellites
\textit{Chandra} and XMM.

\section{Conclusions.}

We inferred the main properties of a TBH in contrast to the widely accepted
SBH model to describe the AGN phenomenon. In particular, we found that a 
rapidly rotating TBH can provide more energy to feed particle acceleration
and production of jets. Besides, a TBH should undergo a transition to a SBH
and so to a quiescent state (if the AGN activity is a pecularity of TBH alone),
forcing nuclear activity to fade and giving a reasonable physical 
explanation to the lifetime of AGNs.\\
The only caveat of our picture is that the model relies on the presence of
a potential V acting as an AdS term, whose origin cannot be due to the
cosmological constant and is missing in the present discussion. The topic
deserves a more strict treatment, anyhow we tend to consider it as a term
due to the surrounding gravitational structure, favouring the TBH collapse.
Other hints could be found in cosmology, for instance in quintessence models
in which the scalar field should have non minimal coupling to gravity, or
else in cosmological models allowing dark matter-dark matter coupling
(e.g. Perrotta, Baccigalupi \& Matarrese 2000, Amendola 2000); nevertheless,
such a cosmological origin is very unlikely, due to the extremely low value
for the potential, preventing its effectiveness in acting as a collapse source.

\section*{Acknowledgements}
The authors are grateful to the Division of High Energy Physics, University
of Helsinki, to the Helsinki Institute of Physics and to the Magnus 
Ehrnrooth Foundation for supporting this work.
The authors wish to thank C. Baccigalupi and F. Perrotta (SISSA), G. Schultz
(Ericsson Research and University of Turku) and M. Vietri (University of Rome 
``RomaTre'') for useful discussions and the anonymous referee for important
comments.


\begin{thebibliography}{10}
\bibitem{} Amendola L. 2000, astro-ph/0006300
\bibitem{} Binney J., Tremaine S. Galactic Dynamics, Princeton
Series in Astrophysics, 1987
\bibitem{} Blandford R.D., Znajek R.L. 1977, MNRAS, 179, 433
\bibitem{} Brill D.R., Louko J., Peldan P. 1997, Phys. Rev. D, 56, 3600
\bibitem{} Cavaliere A., Vittorini V. 2000, ApJ in press, astro-ph/0006194
\bibitem{} Chandrasekhar S. The Mathematical Theory of Black Holes,
Oxford University Press, 1983
\bibitem{} Dendy R., Plasma Physics, Cambridge University Press, 1993
\bibitem{} Fabian A.C., Rees M.J., Stella L., White N.E. 1989, MNRAS, 238, 729
\bibitem{} Frank J., King A., Raine D., Accretion Power in 
Astrophysics, Cambridge Astrophysics Series, 1987
\bibitem{} Hawley J.F., Krolik J.H. 2000, astro-ph/0006456
\bibitem{} Igumenshchev I.V., Abramowicz M.A. 2000, astro-ph/0003397
\bibitem{} Lawden D.F., An Introduction to Tensor Calculus,
Relativity and Cosmology, Wiley \& Sons 1982
\bibitem{} Longair M.S.,High Energy Astrophysics,Cambridge 
University Press, 1981  
\bibitem{} Matt G., Perola G.C., Stella L. 1993, A\&A, 267, 643
\bibitem{} Monaco P., Salucci P., Danese L. 1999, astro-ph/9909267
\bibitem{} Penrose R. 1969, Riv. Nuovo Cimento, 1, Numero Speciale, 252
\bibitem{} Perlmutter S., Gabi S., Goldhaber G. et al. 1997, ApJ, 483, 565
\bibitem{} Perlmutter S., Aldering G., Goldhaber G. et al. 1999, ApJ, 517, 565
\bibitem{} Perrotta F., Baccigalupi C., Matarrese S. 2000, Phys. Rev. D, 61,
023507
\bibitem{} Rees M.J., Phinney E.S., Begelman M.C., Blandford R.D. 1982, 
Nat., 295,17
\bibitem{} Rybicki G., Lightman A. Radiative Processes in Astrophysics, 
Wiley \& Sons, 1979 
\bibitem{} Riess A.G., Filippenko A.G., Challis P. et al. 1998, AJ, 116, 1009
\bibitem{} Roos M., Harun-or-Rashid 2000, astro-ph/0005541
\bibitem{} Smith W.L., Mann R.B. 1997, Phys. Rev. D, 56, 8
\bibitem{} Spivey R.J. 2000, MNRAS, 316, 856
\bibitem{} Vanzo L. 1997,Phys Rev. D, 56, 6475
\bibitem{} Wald R.M., General Relativity, University of Chicago Press,
1984
\end{thebibliography}
\end{document}